\documentclass[12pt]{article}%
\usepackage{amssymb,amsbsy}
\usepackage{amsfonts,amsbsy}
\usepackage{amssymb}
\usepackage{graphicx}
\usepackage{amsfonts}
\usepackage{amsmath}%
\usepackage{url}%

\newcommand{\be}{\begin{equation}}
\newcommand{\ee}{\end{equation}}
\newcommand{\ba}{\begin{array}}
\newcommand{\ea}{\end{array}}
\newcommand{\p}{\partial}

\newcommand{\ds}{\displaystyle}
\def\corank{\mathop{\rm corank}\nolimits}
\def\rank{\mathop{\rm rank}\nolimits}
\def\ddim{\mathop{\rm ddim}\nolimits}
\def\dind{\mathop{\rm dind}\nolimits}

\newtheorem{prop}{Proposition}
\newtheorem{cor}{Corollary}
\newtheorem{remk}{Remark}

\begin{document}
\date{December 21, 2007}
\title%
{\protect\vspace*{-15mm} \bf Generalized St\"ackel Transform\\ and
Reciprocal Transformations\\ for Finite-Dimensional Integrable
Systems}
\author{Artur Sergyeyev$^1$ and Maciej B\l aszak$^2$\\ 
$^1$Mathematical Institute, Silesian University in Opava,\\ Na
Rybn\'\i {}\v{c}ku 1, 746\,01 Opava, Czech Republic\\
$^2$Institute of Physics, A. Mickiewicz University,\\
Umultowska 85, 61-614 Pozna\'{n}, Poland\\
E-mail: {\tt Artur.Sergyeyev@math.slu.cz} and {\tt blaszakm@amu.edu.pl}}
\maketitle

\begin{abstract}\vspace{-10mm}
We present a multiparameter generalization of the St\"ackel
transform (the latter is also known as the coupling-constant metamorphosis) and
show that under certain conditions
this generalized St\"ackel transform
preserves Liouville integrability, noncommutative integrability
and superintegrability.
The corresponding transformation for the equations of motion proves to be
nothing but a reciprocal transformation of a special form, and we
investigate the properties of this reciprocal transformation.
\looseness=-1

Finally, we show that the Hamiltonians of the systems possessing
separation curves of apparently very different form can be related
through a suitably chosen generalized St\"ackel transform.



\medskip

{\bf Keywords:} multiparameter generalized St\"ackel transform,
integrable systems, separation curves, reciprocal transformation

\end{abstract}





\section{Introduction}

The St\"ackel transform \cite{bkm}, also known as the
coupling-constant metamorphosis \cite{hiet}, cf.\ also \cite{kkmw,
kkmw2,kkmw3,ts,ts2} for more recent developments, is a powerful tool
for producing new Liouville integrable systems from the known ones.
This is essentially a transformation that sends an $n$-tuple of
functions in involution on a $2n$-dimensional symplectic manifold into
another $n$-tuple of functions on the same manifold, and these $n$
new functions are again in involution. In its original form the
St\"ackel transform affects just one coupling constant which enters
the Hamiltonian linearly and interchanges this constant with the
energy eigenvalue, see \cite{bkm, hiet}.\looseness=-1

In the
present paper we introduce a multiparameter generalization of the classical
St\"ackel transform, which, just like its known counterpart,
enables us to generate new Liouville
integrable systems from the known ones
or bring known integrable systems into a simpler form.
Unlike the original St\"ackel transform  \cite{bkm, hiet}
this {\em multiparameter generalized St\"ackel
transform} allows for the Hamiltonians being {\em nonlinear} functions
of {\em several} parameters. These properties considerably increase
the power of the transform in question.
\looseness=-1

Most importantly, under certain natural assumptions
the multiparameter generalized St\"ackel transform
preserves Liouville integrability, superintegrability
and noncommutative integrability, see Propositions~\ref{trp}~and~\ref{nci}
and the discussion thereafter.

Moreover, in Section~\ref{rtem} we show that the transformations for equations
of motion induced by the multiparameter generalized St\"ackel transform
are nothing but reciprocal transformations. This
generalizes to the multiparameter case the earlier results of
Hietarinta et al.\ \cite{hiet} on the one-parameter St\"ackel
transform. \looseness=-1


The significance of reciprocal transformations in the theory of
integrable nonlinear partial differential equations is well
recognized. These transformations were intensively used in the
theory of dispersionless (hydrodynamic-type) systems as well as in
the theory of soliton systems, see e.g.\ \cite{cr2,cr1} and
references therein. \looseness=-1
On the other hand, some particular examples of
transformations of this kind for
finite-dimensional Hamiltonian systems are also known, for instance
the Jacobi transformation, see \cite{lan}
and a recent survey \cite{ts2}. The reciprocal transformations of
somewhat different kind have also appeared in \cite{hiet, ves,
ts}.

In the present paper we consider reciprocal transformations for
the Liouville integrable Hamiltonian systems in conjunction with the
generalized St\"ackel transform and, in contrast with the earlier
work on the subject, we concentrate on the multi-time version of
these transformations.



In fact, as we show in Section~\ref{rtem}
below, these transformations, when applied to the equations
of motion of the source system, in general do \emph{not} yield the equations
of motion for the target system {\em unless} we restrict the equations of motion
onto the common level surface of the corresponding Hamiltonians, see
Propositions~\ref{eomp}~and~\ref{eomp1} below for details.

We further show that for two Liouville integrable systems related by an
appropriate multiparameter generalized St\"ackel transform for the constants of motion
we have the reciprocal transformation relating the corresponding equations of motion
restricted to appropriate Lagrangian submanifolds, see e.g.\ Ch.3 of
\cite{cas} and references therein
for more details on the latter.

Moreover, we present a multitime extension of the original reciprocal
transformation from 
\cite{hiet}, and study the applications
of this extended transformation to the integration of equations of motion in
the Hamilton--Jacobi formalism using the separation of variables, cf.\ \cite{bkm}.





In the rest of the paper we consider the relations among
classical Liouville integrable St\"ackel systems on
$2n$-dimensional phase space. In \cite{mac2005} infinitely many
classes of the St\"ackel systems related to the so-called seed
class,
namely, the $k$-hole deformations of the latter, were constructed.
Here we show that any $k$-hole deformation can be obtained from
the Benenti-type system through a
suitably chosen multiparameter generalized
St\"ackel transform, and present the explicit form of the
transform in question along with its inverse.
\looseness=-1

\section{Multiparameter generalized St\"ackel transform:\\
definition and duality}\label{mgstdefsect}
Let $(M,P)$ be a Poisson manifold with the Poisson
bracket $\{f,g\}=(df, Pdg)$. Consider $r$ functionally independent
Hamiltonians $H_i$, $i=1,\dots,r$, on $M$,
and assume that these Hamiltonians
further depend on $k\leq r$ parameters $\alpha_1,\dots,\alpha_k$, so
\begin{equation}
\label{df}H_{i}=H_{i}(x,\alpha_1,\dots,\alpha_k),\quad i=1,\dots,r,
\end{equation}
where $x\in M$. Note that in general $r$ is not related in any way
to the dimension of $M$ except for the obvious restriction
$r\leq\dim M$; see, however, the discussion after
Proposition~\ref{trp}. Also, in what follows all functions
will be tacitly assumed to be smooth (of the $C^\infty$ class).

Suppose that there exists
a $k$-tuple of pairwise distinct numbers $s_i\in\{1,\dots,r\}$
such that
\begin{equation}
\label{det}\det\left(|\!|\p
H_{s_i}/\p\alpha_j|\!|_{i,j=1,\dots,k}\right)\neq 0.
\end{equation}

Now fix a $k$-tuple $\{s_1,\dots,s_k\}$ such that (\ref{det}) holds
and consider the system
\[
H_{s_i}(x,\alpha_1,\dots,\alpha_k)=\tilde{\alpha}_i,\quad i=1,\dots,k,
\]
where $\tilde \alpha_i$ are arbitrary parameters,
as a system of algebraic equations for $\alpha_1,\dots,\alpha_k$.
By the implicit function theorem, the condition (\ref{det})
guarantees that the solution of this system exists and
is (locally) unique. We can write this solution in the form
\[
\alpha_i=A_i(x,\tilde\alpha_1,\dots,\tilde\alpha_k),\quad i=1,\dots,k.
\]
Now define the new Hamiltonians $\tilde H_{s_i}$, $i=1,\dots,k$, by setting
\[
\tilde H_{s_i}=A_i(x,\tilde\alpha_1,\dots,\tilde\alpha_k), \quad i=1,\dots,k.
\]

In other words, the Hamiltonians $\tilde H_{s_i}$, $i=1,\dots,k$
are defined by
means of the relations
\begin{equation}
\label{dual1} H_{s_i}|_{[\Phi]}=\tilde{\alpha}_i,\quad i=1,\dots,k.
\end{equation}
Here and below the subscript $[\Phi]$
means that we have substituted $\tilde H_{s_i}$
for $\alpha_i$ for all $i=1,\dots,k$.

Next, let
\begin{equation}
\label{hi}%
\tilde H_{i}
=H_{i}|_{[\Phi]}, \quad i=1,\dots,r, \quad i\neq
s_j\quad\mbox{for}\quad\! j=1,\dots,k.
\end{equation}

Note that the Hamiltonians $\tilde H_j$
involve $k$ parameters $\tilde\alpha_i$, $i=1,\dots,k$ for all $j=1,\dots,r$:
\[
\tilde H_i=\tilde H_i(x,\tilde\alpha_1,\dots,\tilde\alpha_k),\quad i=1,\dots,r.
\]

We shall refer to the above transformation from $H_i$,
$i=1,\dots,r$, to $\tilde H_i$, $i=1,\dots,r$, as to the
$k$-parameter {\em generalized St\"ackel transform} generated by
$H_{s_1},\dots, H_{s_k}$. In analogy with \cite{bkm} we shall say
that the $r$-tuples $H_i$, $i=1,\dots,r$, and $\tilde H_i$,
$i=1,\dots,r$, are {\em St\"ackel-equivalent}.

The condition (\ref{det}) guarantees that the above transformation is invertible.
Indeed, consider the
dual of the identity (\ref{dual1}), that is,
\begin{equation}
\label{dual2} \tilde H_{s_i}|_{[\tilde\Phi]}=\alpha_i,\quad i=1,\dots,k,\quad
\end{equation}
where the subscript $[\tilde\Phi]$
means that we have substituted $H_{s_i}$
for $\tilde{\alpha}_i$ for all $i=1,\dots,k$.

Moreover, the functional independence of the original Hamiltonians
$H_i$, $i=1,\dots,r$, implies the functional independence of $\tilde
H_i$, $i=1,\dots,r$. Indeed, the functional independence of
$H_i$, $i=1,\dots,r$, means that $\dim\mathrm{span} (d
H_i, i=1,\dots,r)=r$ on another open dense subset $U$ of $M$.
Using (\ref{det}), (\ref{dual1}) and (\ref{hi}) we readily see
that this implies $\dim\mathrm{span} (d
H_i, i=1,\dots,r)=r$ on another open dense subset $\tilde U\subset U$ of $M$.
In turn, the latter equality means nothing but the functional
independence of $\tilde H_i$, $i=1,\dots,r$ we sought for.

Let us stress that here and below the differentials are computed under
the assumption that the parameters are considered to be constant,
i.e., if $H=H(x,\alpha_1,\dots,\alpha_k)$ then in the local
coordinates $x^b$ on $M$ we have\looseness=-1
\[
d H=\sum\limits_{b=1}^{\dim M}\displaystyle\frac{\partial
H}{\partial x^b} dx^b.
\]
\looseness=-1

By the implicit function theorem the condition (\ref{det})
guarantees that we can solve (\ref{dual2}) with respect to $H_{s_j}$, $j=1,\dots,k$.
If we do this and define the remaining Hamiltonians $H_i$ by the formulas
\begin{equation}
\label{hii}
H_{i}
=\tilde H_{i}|_{[\tilde\Phi]}, \quad i=1,\dots,r, \quad i\neq
s_j\quad\mbox{for}\quad\! j=1,\dots,k,
\end{equation}
then it is straightforward to verify
that (\ref{dual1}) and (\ref{hi}) hold identically. In other words, the formulas (\ref{dual2}) and
(\ref{hii}) define the inverse of the transformation defined using (\ref{dual1}) and (\ref{hi}).

Clearly, these two transformations are dual,
with the duality transformation swapping $H_i$ and $\tilde H_i$ for all $i=1,\dots,r$
and swapping $\alpha_j$ and $\tilde\alpha_j$ for all $j=1,\dots,k$.

Note that in the special case when the Hamiltonians
$H_i$ are linear in the parameters $\alpha_j$,
the above formulas undergo considerable simplification,
and we can explicitly  express $\tilde H_i$ via $H_i$.

Namely, let
\begin{equation}
\label{dflin}H_{i}=H_{i}^{(0)} +\sum\limits_{j=1}^k \alpha_j
H_{i}^{(j)},\quad i=1,\dots,r.
\end{equation}
Then equations (\ref{dual1}) take the form
\begin{equation}
\label{dual1lin0} H_{s_i}^{(0)}+\sum\limits_{j=1}^k \tilde H_{s_j}
H_{s_i}^{(j)}=\tilde{\alpha}_i,\quad i=1,\dots,k,
\end{equation}
and we can readily solve them for $\tilde H_{s_i}$:
\begin{equation}\label{dual1lin}
\tilde H_{s_i}=\det W_i/\det W,
\end{equation}
where $W$ is a $k\times k$ matrix of the form
\[
W=\left|\!\left|\begin{array}{ccc}
H_{s_1}^{(1)} & \cdots & H_{s_1}^{(k)}\\
\vdots & \ddots & \vdots\\
H_{s_k}^{(1)} & \cdots & H_{s_k}^{(k)}
\end{array}\right|\!\right|,
\]
and $W_i$ are obtained from $W$ by replacing
$H_{s_j}^{(i)}$ by $H_{s_j}^{(0)}-\tilde\alpha_j$ for all
$j=1,\dots,k$.

By (\ref{hi}) we have
\begin{equation}
\label{hilin}%
\tilde H_{i}
=H_{i}^{(0)} +\sum\limits_{j=1}^k \tilde H_{s_j} H_{i}^{(j)},
\qquad i=1,\dots,r, \quad i\neq s_j\quad\mbox{for}\quad\!
j=1,\dots,k,
\end{equation}
where $\tilde H_{s_i}$ are given by (\ref{dual1lin}).
It is straightforward to verify that if we set $k=1$
then the
transformation given by (\ref{dual1lin}) and (\ref{hilin})
becomes nothing but the standard St\"ackel transform
\cite{bkm}, also known as the coupling-constant metamorphosis \cite{hiet}.

\section{Multiparameter generalized St\"ackel transform\\
and (super)in\-tegrability}\label{mgstsov}

It turns out that the $k$-parametric generalized St\"ackel transform
preserves the commutativity of the Hamiltonians $H_i$.
More precisely, we have the following result:
\begin{prop}
\label{trp} 
Let $H_i$, $i=1,\dots,r$, be functionally independent
and let $\tilde H_i$, $i=1,\dots,r$, be related to $H_i$, $i=1,\dots,r$,
by a $k$-parameter generalized St\"ackel transform (\ref{dual1}), (\ref{hi}) generated by $H_{s_1},\dots, H_{s_k}$,
where $k\leq \corank P+(1/2)\rank P$.

Then the following assertions hold:
\begin{itemize}
\item[i)] if $\{H_{s_i}, H_{s_j}\}=0$
for all $i,j=1,\dots,k$ then $\{\tilde H_{s_i}, \tilde H_{s_j}\}=0$
for all $i,j=1,\dots,k$;  
\item[ii)] suppose that i) holds and
for a $j_0\in\{1,\dots,r\}$, $j_0\neq s_1,\dots,s_k$, we have
$\{H_{s_i}, H_{j_0}\}=0$
for all $i=1,\dots,k$; then $\{\tilde
H_{s_i}, \tilde H_{j_0}\}=0$ for all $i=1,\dots,k$;
\item[iii)] suppose that for a natural $m\leq \corank P+(1/2)\rank P$
we have an $m$-tuple of pairwise distinct
integers $l_1,\dots,l_m\in\{1,\dots,r\}$
such that $s_q\in\{l_1,\dots,l_m\}$ for all $q=1,\dots,k$,
and $\{H_{l_i}, H_{l_j}\}=0$ for all $i,j=1,\dots,m$;
then $\{\tilde H_{l_i}, \tilde H_{l_j}\}=0$
for all $i,j=1,\dots,m$.
\end{itemize}
\end{prop}

Before we proceed with the proof of Proposition~\ref{trp}, some
remarks are in order. First of all, $\corank P+(1/2)\rank P$ is easily seen to be
the maximal possible number of functions in involution on $M$ with
respect to the Poisson bracket associated with $P$.

Next, from Proposition~\ref{trp} it is immediate that the transformation
defined by (\ref{dual1}) and (\ref{hi}) preserves
(super)integrability.
Namely, under the assumptions of Proposition~\ref{trp}, iii) let
$\dim M=2n$, $\mathop{\rm rank}P=2n$, and $m=n$. Then the dynamical
system associated with any of $H_{l_i}$ is Liouville integrable, as it
has $n$ commuting functionally independent integrals, $H_{l_j}$,
$j=1,\dots,n$, in involution. By Proposition~\ref{trp}, iii) the
dynamical system associated with any of $\tilde H_{l_i}$ enjoys the same
property, the required integrals of motion in involution now being
$\tilde H_{l_i}$, $i=1,\dots,n$.

Note that if under the assumptions of Proposition~\ref{trp}, iii) we have
$\dim M=2n$, $\mathop{\rm rank}P=2n$, and $m<n$ then, under some
technical assumptions and in a suitable vicinity $U\subset M$,
for the dynamical system associated with any
of $H_{l_i}$, $i=1,\dots,m$,  there
exists a symplectic submanifold fibred into $m$-dimensional
invariant tori \cite{nekh72, nekh, ga}. The tori in question
are intersections of this symplectic submanifold with the
common level surfaces of $H_{l_i}$, $i=1,\dots,m$.
Proposition~\ref{trp}, iii) implies that
this property is preserved by the multiparameter
generalized St\"ackel transform defined by (\ref{dual1}) and
(\ref{hi}), i.e., for the dynamical system associated with any of
$\tilde H_{l_i}$, $i=1,\dots,m$, there exists, again under certain
technical assumptions and in a suitable vicinity $\tilde U\subset M$,
a symplectic submanifold fibred into
$m$-dimensional invariant tori.\looseness=-1

Now let $\dim M=2n$, $\mathop{\rm rank}P=2n$, $r>n$, and suppose that
$\{H_{s_i},H_j\}=0$ for all $i=1,\dots,k$ and for all $j=1,\dots,r$. Then the
Hamiltonian $H_{s_j}$ is superintegrable for any $j\in\{1,\dots,k\}$
as it has $r>n$ integrals of motion $H_{i}$, $i=1,\dots,r$, and by
Proposition~\ref{trp}, ii) the
Hamiltonian $\tilde H_{s_j}$ is superintegrable for
any $j\in\{1,\dots,k\}$ as well, the integrals of motion now being $\tilde
H_i$, $i=1,\dots,r$. \looseness=-1


Moreover, the multiparameter generalized
St\"ackel transform defined by (\ref{dual1}) and (\ref{hi}) also
preserves noncommutative integrability in the sense of \cite{mif, bj}.
We start with the following result:

\begin{prop}\label{nci}
Under the assumptions of Proposition~\ref{trp}, iii)
suppose that $\dim M=2n$, $P$ is nondegenerate ($\rank P=2n$), and
the algebra $\mathcal{F}$
of functions on $M$ generated by $H_1,\dots,H_r$ is closed under
the Poisson bracket and is complete in the sense of \cite{bj}.
Further suppose that $\ker \{,\}|_{\mathcal{F}}=\mathcal{F}_0$,
where $\mathcal{F}_0$ is the algebra
of functions on $M$ generated by $H_{l_1},\dots,H_{l_m}$.

Then the algebra $\tilde{\mathcal{F}}$
of functions on $M$ generated by $\tilde{H}_1,\dots,\tilde{H}_r$
is also closed under the Poisson bracket and
complete. 
\end{prop}

Consider an algebra
$\mathcal{A}$ of functions on a {\em symplectic} manifold $M$ and assume that
$\mathcal{A}$ is closed under the Poisson bracket. Recall (see \cite{bj} for precise definitions and
further details) that the {\em differential dimension} $\ddim\mathcal{A}$ of $\mathcal{A}$
is, roughly speaking, the number of functionally independent generators of
$\mathcal{A}$. The {\em differential index} $\dind\mathcal{A}$ can be (informally)
defined as $\dind\mathcal{A}=\ddim\ker \{,\}|_{\mathcal{A}}$, and $\mathcal{A}$
is said to be {\em complete} \cite{bj}
if $\ddim{\mathcal{A}}+\dind{\mathcal{A}}=\dim M$
on an open dense subset $U\subset M$.\looseness=-1

{\em Sketch of proof of Proposition~\ref{nci}.} First of all,
it is immediate that the algebra $\tilde{\mathcal{F}}$
generated by $\tilde{H}_1,\dots,\tilde{H}_r$ is also closed under
the Poisson bracket.
As we have already noticed in Section~\ref{mgstdefsect},
the functional independence of $H_i$, $i=1,\dots,r$, implies that
of $\tilde H_i$, $i=1,\dots,r$, and hence we have
$\ddim\tilde{\mathcal{F}}=\ddim\mathcal{F}=r$.
In turn, as $\ker \{,\}|_{\mathcal{F}}=\mathcal{F}_0$, we have
$\dind\mathcal{F}=\ddim\mathcal{F}_0=m$.

By Proposition~\ref{trp}, iii)
we have $\{\tilde H_{l_i},\tilde H_{l_j}\}=0$, $i,j=1,\dots,m$,
so $\ker \{,\}|_{\tilde{\mathcal{F}}}\supset\tilde{\mathcal{F}}_0$,
where $\tilde{\mathcal{F}}_0$ is the algebra of functions on $M$
generated by $\tilde{H}_{l_1},\dots,\tilde{H}_{l_m}$. Therefore
$\dind\tilde{\mathcal{F}}\geq \ddim\tilde{\mathcal{F}}_0=m$. However, as we obviously have
$\ddim\tilde{\mathcal{F}}+\dind\tilde{\mathcal{F}}\leq \dim M$
and, on the other hand, we know from the above that
$\ddim\tilde{\mathcal{F}}+\dind\tilde{\mathcal{F}}\geq r+m=\dim M$,
we conclude that $\ddim\tilde{\mathcal{F}}+\dind\tilde{\mathcal{F}}=\dim M$,
and thus the algebra $\tilde{\mathcal{F}}$ is indeed complete. $\square$

Therefore, if under the assumptions of Proposition~\ref{nci}
there exists an integer $i_0\in\{1,\dots,r\}$ such that
$\{H_{i_0},H_j\}~=~0$, $j=1,\dots,r$,
and thus the dynamical system associated with $H_{i_0}$
is completely integrable in the noncommutative sense \cite{mif, bj}, as this system
possesses a complete algebra of integrals of motion, then
so does the dynamical system associated with
$\tilde H_{i_0}$.

{\em Proof of Proposition~\ref{trp}.} Prove i) first. For any smooth
functions $f$ and $g$ on $M$ that further depend on the parameters
$\alpha_1,\dots,\alpha_k$, we have the following easy identities:
\begin{equation}
\left.\{f|_{[\Phi]},g\}\right|_{[\Phi]}=\left.\{f,g\}\right|_{[\Phi]}
+\sum\limits_{j=1}^k (\p f/\p\alpha_j)|_{[\Phi]} \{\tilde H_{s_j}, g\}|_{[\Phi]},\label{pb1}
\end{equation}
\begin{equation}\begin{array}{r}
\{f|_{[\Phi]},g|_{[\Phi]}\}=\left.\{f,g\}\right|_{[\Phi]}
+\sum\limits_{j=1}^k (\p f/\p\alpha_j)|_{[\Phi]} \{\tilde H_{s_j}, g\}|_{[\Phi]}
+\sum\limits_{j=1}^k (\p g/\p\alpha_j)|_{[\Phi]} \{f, \tilde H_{s_j}\}|_{[\Phi]}\\
+\sum\limits_{i,j=1}^k (\p f/\p\alpha_i)|_{[\Phi]} (\p
g/\p\alpha_j)|_{[\Phi]} \{\tilde H_{s_i}, \tilde
H_{s_j}\}.\end{array}\label{pb2}
\end{equation}

Using the assumption $\{H_{s_i},H_{s_j}\}=0$ and (\ref{dual1}),
we find that 
\[
0=\{\tilde{\alpha}_i-H_{s_i},
\tilde{\alpha}_j-H_{s_j}\}=
\{H_{s_i}|_{[\Phi]}-H_{s_i},
H_{s_j}|_{[\Phi]}-H_{s_j}\},
\]
whence
\[
\{H_{s_i}|_{[\Phi]}-H_{s_i}, H_{s_j}|_{[\Phi]}-H_{s_j}\}|_{[\Phi]}=0.
\]
Writing out the Poisson bracket
on the left-hand side of the latter identity
using (\ref{pb1}) for the brackets
$\{H_{s_i}|_{[\Phi]},H_{s_j}\}|_{[\Phi]}$ and
$\{H_{s_i}, H_{s_j}|_{[\Phi]}\}|_{[\Phi]}$ and (\ref{pb2}) for the bracket
$\{H_{s_i}|_{[\Phi]},H_{s_j}|_{[\Phi]}\}$
we obtain
\[
\sum\limits_{p,q=1}^k (\p H_{s_i}/\p\alpha_p)|_{[\Phi]} (\p
H_{s_j}/\p\alpha_q)|_{[\Phi]} \{\tilde H_{s_p}, \tilde H_{s_q}\}|_{[\Phi]}=0,
\]
whence using (\ref{det}) we readily find that for all
$p,q=1,\dots,k$ we have
\[
 \{\tilde H_{s_p}, \tilde H_{s_q}\}|_{[\Phi]}=0.
\]

However, $\tilde H_{s_p}$ are independent of $\alpha_q$
for all $q=1,\dots,k$, so
\[
\{\tilde H_{s_i}, \tilde H_{s_j}\}=\{\tilde H_{s_i}, \tilde H_{s_j}\}|_{[\Phi]}=0,
\]
and the result follows.

As we have already proved i), to prove ii) we only need to show that
if $\{H_{s_i}, H_{j_0}\}=0$
for all $i=1,\dots,k$ then
$\{\tilde H_{s_i}, \tilde H_{j_0}\}=0$
for all $i=1,\dots,k$.

As $\tilde H_i$, $i=1,\dots,r$, are independent of $\alpha_p$
for all $p=1,\dots,k$ by construction, we have
\[
\{\tilde H_{s_i}, \tilde H_{j_0}\}=\{\tilde H_{s_i}, \tilde H_{j_0}\}|_{[\Phi]}.
\]

Moreover, as $j_0\neq s_p$ for all $p=1,\dots,k$ by assumption, by virtue of
(\ref{hi}) the relation
$\{\tilde H_{s_i}, \tilde H_{j_0}\}|_{[\Phi]}=\nobreak 0$ is equivalent to
\[
\{\tilde H_{s_i}, H_{j_0}|_{[\Phi]}\}|_{[\Phi]}=0.
\]
In turn, using (\ref{pb1}) we can rewrite the Poisson bracket $\{\tilde H_{s_i},
H_{j_0}|_{[\Phi]}\}|_{[\Phi]}$ as follows:
\[
\{\tilde H_{s_i}, H_{j_0}|_{[\Phi]}\}|_{[\Phi]}=\{\tilde H_{s_i},
H_{j_0}\}|_{[\Phi]} -\sum\limits_{p=1}^k (\p
H_{j_0}/\p\alpha_p)|_{[\Phi]} \left.\{\tilde H_{s_p}, \tilde
H_{s_i}\}\right|_{[\Phi]}
\]
As 
$\{\tilde H_{s_p}, \tilde H_{s_i}\}=0$ by i),
we see that
\[
\{\tilde H_{s_i}, H_{j_0}|_{[\Phi]}\}|_{[\Phi]}=\{\tilde H_{s_i},
H_{j_0}\}|_{[\Phi]}.
\]
Now, in analogy with the proof of i), consider the identity
\[
0=\{\tilde\alpha_p,H_{j_0}\}|_{[\Phi]}
=\{H_{s_p}|_{[\Phi]},H_{j_0}\}|_{[\Phi]}.
\]
Using (\ref{pb1}) and our assumptions yields
\[
0=\{H_{s_p}|_{[\Phi]},H_{j_0}\}|_{[\Phi]}
=\sum\limits_{i=1}^k (\p H_{s_p}/\p\alpha_i)|_{[\Phi]}\{\tilde H_{s_i},H_{j_0}\}|_{[\Phi]}.
\]
Finally, using (\ref{det}) we conclude that
\begin{equation}\label{pb3}\{\tilde H_{s_i},H_{j_0}\}|_{[\Phi]}=0,
\end{equation}
whence $\{\tilde H_{s_i}, H_{j_0}|_{[\Phi]}\}|_{[\Phi]}=0$,
and the result follows.

Part iii) is proved in analogy with ii). Namely,
in view of i) and ii) we only need to prove that
the conditions $\{H_{l_i},H_{l_j}\}=0$, $i,j=1,\dots,m$ imply
$\{\tilde H_{l_i}, \tilde H_{l_j}\}=0$
for all $i,j=1,\dots,m$ such that $l_i\neq s_p$ and
$l_j\neq s_p$ for all $p=1,\dots,k$.

If $l_i\neq s_p$ and
$l_j\neq s_p$ for all $p=1,\dots,k$ then we have
\[
\{\tilde H_{l_i}, \tilde H_{l_j}\}
=\{H_{l_i}|_{[\Phi]}, H_{l_j}|_{[\Phi]}\}.
\]
Using (\ref{pb2}) and (\ref{pb3}) for $j_0=l_i$ and $j_0=l_j$ we readily find that
\[
\{H_{l_i}|_{[\Phi]}, H_{l_j}|_{[\Phi]}\}=0,
\]
and the result follows.
$\square$

Note that the computations in the above proof
bear considerable resemblance to those in the theory of
Hamiltonian systems with second-class constraints, see e.g.\ the
classical book of Dirac \cite{dirac}.

\section{Reciprocal transformations for the equations of motion}\label{rtem}
Recall that the equations of motion associated with a Hamiltonian $H$ and a
Poisson structure $P$ on $M$ read (see e.g.\ \cite{mb})
\begin{equation}
\label{eom0}d x^{b}/dt_{H} =(X_{H})^{b},\quad b=1,\dots,\dim M,
\end{equation}
where $x^{b}$ are local coordinates on $M$, $X_{H}=P dH$ is the
Hamiltonian vector field associated with $H$, and $t_{H}$ is the
corresponding evolution parameter (time).

Throughout the rest of this section we tacitly assume that $\tilde H_i$, $i=1,\dots,r$,
are related to $H_i$, $i=1,\dots,r$, through the $k$-parameter St\"ackel transform
(\ref{dual1}), (\ref{hi}) generated by $H_{s_1},\dots, H_{s_k}$.

Suppose that $\{H_{s_i}, H_{s_j}\}=0$
for all $i,j=1,\dots,k$ ,and consider
simultaneously the equations of motion (\ref{eom0}) for the Hamiltonians
$H_{s_i}$ with the times $t_{s_i}$ and
for $\tilde H_{s_i}$ with the times $\tilde t_{s_i}$:
\begin{align}
d x^{b}/dt_{s_i} =(X_{H_{s_i}})^{b},\quad b=1,\dots,\dim M,\quad i=1,\dots,k,\label{eom1}\\[3mm]
d x^{b}/d\tilde t_{s_i} =(X_{\tilde H_{s_i}})^{b},\quad b=1,\dots,\dim M,\quad i=1,\dots,k. \label{eom2}%
\end{align}

In analogy with \cite{hiet} consider a \emph{reciprocal
transformation} (see e.g.\ \cite{cr2,cr1,rj} for general information
on such transformations) relating the times $t_{s_i}$ and $\tilde
t_{s_j}$:
\begin{equation}
\label{rt0}d\tilde t_{s_i}=-\sum\limits_{j=1}^k \left.\left(\frac{\p
H_{s_j}}{\p\alpha_i}\right) \right|_{[\Phi]} dt_{s_j},\quad
i=1,\dots,k.
\end{equation}
\begin{prop}
\label{eomp} 
Suppose that $k\leq \corank P+(1/2) \rank P$
and $\{H_{s_i}, H_{s_j}\}=0$
for all $i,j=1,\dots,k$, and
consider the equations of motion (\ref{eom1}) for $H_{s_i}$,
$i=1,\dots,k$, restricted onto the common level surface $N_{\tilde\alpha}$
of $H_{s_i}$, where
\[
N_{\tilde\alpha}=\{x\in M|H_{s_i}(x,\alpha_1,\dots,\alpha_k)
=\tilde\alpha_i,\quad i=1,\dots,k\}.
\]

Then the transformation (\ref{rt0}) is well defined on these
restricted equations of motion and sends them into the equations of
motion (\ref{eom2}) for $\tilde H_{s_i}$, $i=1,\dots,k$, restricted
onto the common level surface $\tilde N_\alpha$ of $\tilde H_{s_i}$, where
\[
\tilde N_\alpha=\{x\in M|\tilde H_{s_i}
(x,\tilde{\alpha}_1,\dots,\tilde{\alpha}_k)=\alpha_i,\quad
i=1,\dots,k\}.
\]
\end{prop}
Note that 
the level surfaces in question, $\tilde N_\alpha$ and $N_{\tilde\alpha}$,
represent the {\em same} submanifold of $M$, i.e.,
$\tilde N_\alpha=N_{\tilde\alpha}$.
This is readily verified using the relations (\ref{dual1}) and (\ref{dual2}).

{\em Proof.} First of all show that (\ref{rt0}) is well-defined,
that is, we have
\begin{equation}\label{mixd1}
\frac{\p^2 \tilde t_{s_i}}{\p t_{s_{p}}\p t_{s_q}} =\frac{\p^2
\tilde t_{s_i}}{\p t_{s_{q}}\p t_{s_p}},\quad p,q=1,\dots,k,
\end{equation}
by virtue of equations (\ref{eom1}) restricted onto
$N_{\tilde\alpha}$.

Using (\ref{rt0}) we find that (\ref{mixd1}) boils down to
\begin{equation}\label{mixd1a}
\left.\left(\ds\frac{\p\left.\left(\ds\frac{\p H_{s_p}}{\p\alpha_i}\right)
\right|_{[\Phi]}}{\p t_{s_q}}\right)\right|_{N_{\tilde\alpha}}
=\left.\left(\ds\frac{\p\left.\left(\ds\frac{\p
H_{s_q}}{\p\alpha_i}\right) \right|_{[\Phi]}}{\p
t_{s_p}}\right)\right|_{N_{\tilde\alpha}},\quad p,q=1,\dots,k.
\end{equation}
In turn, using (\ref{eom1}) we readily find that (\ref{mixd1a}) takes the
form
\[
\left.\left\{\left.\left(\frac{\p H_{s_p}}{\p\alpha_i}\right)
\right|_{[\Phi]}, H_{s_q}\right\}\right|_{N_{\tilde\alpha}}
=\left.\left\{\left.\left(\frac{\p H_{s_q}}{\p\alpha_i}\right)
\right|_{[\Phi]}, H_{s_p}\right\}\right|_{N_{\tilde\alpha}},
\]
and the latter equality can be proved by taking the partial derivative of
the relation
$\{H_{s_p},H_{s_q}\}=0$ with respect
to $\alpha_i$. 

Next, Eq.(\ref{rt0}) yields
\[
\ds\frac{d}{dt_{s_i}}=-\sum\limits_{j=1}^k \left.\left(\frac{\p
H_{s_i}}{\p\alpha_j}\right) \right|_{[\Phi]} \ds\frac{d}{d\tilde t_{s_j}},\quad
i=1,\dots,k.
\]
Taking into account (\ref{eom1}) and (\ref{eom2}) we conclude that we
have
to prove that 
\begin{equation}
\label{td0}X_{H_{s_i}}|_{N_{\tilde\alpha}}=-\sum\limits_{j=1}^k
\left.\left(\left.\left(\frac{\p H_{s_i}}{\p\alpha_j}\right)
\right|_{[\Phi]}\right)\right|_{N_{\tilde\alpha}} X_{\tilde
H_{s_j}}|_{N_{\tilde\alpha}},\quad i=1,\dots,k,
\end{equation}
where $|_{N_{\tilde\alpha}}$ denotes restriction onto
$N_{\tilde\alpha}$.

As $X_H=P dH$ for any smooth function $H$ on $M$, Eq.(\ref{td0})
boils down to
\begin{equation}
\label{td0a}\left.\left(P\left(d H_{s_i}+ \sum\limits_{j=1}^k
\left.\left(\frac{\p H_{s_i}}{\p\alpha_j}\right) \right|_{[\Phi]}
d{\tilde H_{s_j}}\right)\right)\right|_{N_{\tilde\alpha}}=0, \quad
i=1,\dots,k.
\end{equation}

On the other hand, taking the differential of (\ref{dual1}) we obtain
\begin{equation}\label{dif}
(d H_{s_i})|_{[\Phi]}+ \sum\limits_{j=1}^k \left.\left(\frac{\p
H_{s_i}}{\p\alpha_j}\right) \right|_{[\Phi]} (d\tilde
H_{s_j})|_{[\Phi]}=0,\quad i=1,\dots,k.
\end{equation}

As $\tilde H_{s_j}$ are independent of $\alpha_p$, for all
$p=1,\dots,k$ we have $(d\tilde H_{s_j})|_{[\Phi]}=d\tilde H_{s_j}$,
so (\ref{dif}) yields
\[
\sum\limits_{j=1}^k \left.\left(\frac{\p H_{s_i}}{\p\alpha_j}\right)
\right|_{[\Phi]} d\tilde H_{s_j}=-(d H_{s_i})|_{[\Phi]},
\]
and (\ref{td0}) takes the form
\[
\left.\left(P\left(d H_{s_i}-(d
H_{s_i})|_{[\Phi]}\right)\right)\right|_{N_{\tilde\alpha}}
=0, \quad
i=1,\dots,k.
\]

In the local coordinates $x^b$ on $M$ we have  
\begin{equation}\label{td0b}\begin{array}{l} \ \left.\left(P\left(d H_{s_i}-(d
H_{s_i})|_{[\Phi]}\right)\right)\right|_{N_{\tilde\alpha}}
=\left.\left(P\left(\sum\limits_{b=1}^{\dim
M}\left(\displaystyle\frac{\p H_{s_i}}{\p x^b}-\left.\left(\frac{\p
H_{s_i}}{\p x^b}\right)\right|_{[\Phi]}
\right)dx^b\right)\right)\right|_{N_{\tilde\alpha}}\\[3mm]
=\sum\limits_{b=1}^{\dim M}\left.\left(\displaystyle\frac{\p
H_{s_i}}{\p x^b}-\left.\left(\frac{\p H_{s_i}}{\p
x^b}\right)\right|_{[\Phi]}
\right)\right|_{N_{\tilde\alpha}}\left.\left(P
dx^b\right)\right|_{N_{\tilde\alpha}}, \quad i=1,\dots,k.
\end{array}
\end{equation}

By virtue of (\ref{dual1}) and (\ref{dual2}) $N_{\tilde\alpha}$
and $\tilde N_{\alpha}$ represent the same submanifold of $M$,
whence
\[
\left.\left(\displaystyle\frac{\p H_{s_i}}{\p
x^b}-\left.\left(\frac{\p H_{s_i}}{\p x^b}\right)\right|_{[\Phi]}
\right)\right|_{N_{\tilde\alpha}}=\left.\left(\displaystyle\frac{\p
H_{s_i}}{\p x^b}-\left.\left(\frac{\p H_{s_i}}{\p
x^b}\right)\right|_{[\Phi]} \right)\right|_{\tilde
N_{\alpha}}=\left.\left(\displaystyle\frac{\p H_{s_i}}{\p
x^b}\right)\right|_{\tilde N_{\alpha}}-\left.\left(\frac{\p
H_{s_i}}{\p x^b}\right)\right|_{\tilde N_{\alpha}}=0.
\]
We used here 
an easy identity
\[
\left.\left(\left.\left(\frac{\p H_{s_i}}{\p
x^b}\right)\right|_{[\Phi]} \right)\right|_{\tilde N_{\alpha}}=
\left.\left(\frac{\p H_{s_i}}{\p x^b} \right)\right|_{\tilde
N_{\alpha}}.
\]
\nopagebreak[4]
Thus, the left-hand side of (\ref{td0b}), and therefore that of (\ref{td0a}),
vanishes,
and the result follows.
$\square$

Now assume that all $H_{i}$ are in involution:
\[
\{H_{i},H_{j}\}=0,\quad i,j=1,\dots,r.
\]
Then by Proposition \ref{trp},iii) so are $\tilde H_{i}$, i.e.,
\[
\{\tilde H_{i},\tilde H_{j}\}=0,\quad i,j=1,\dots,r,
\]
and we can consider two sets of simultaneous evolutions,
\begin{align}
d x^{b}/dt_{i} =(X_{H_{i}})^{b},\quad b=1,\dots,\dim M,\quad i=1,\dots
,r,\label{eomk1}\\[3mm]
d x^{b}/d\tilde t_{i} =(X_{\tilde H_{i}})^{b},\quad b=1,\dots,\dim M,\quad
i=1,\dots,r, \label{eomk2}%
\end{align}
and the following extension of (\ref{rt0}):
\begin{equation}
\begin{array}{l}
\label{rct}\displaystyle
d\tilde t_{s_i}=-\sum\limits_{j=1}^r \left.\left(\frac{\p H_{j}}{\p\alpha_i}\right)
\right|_{[\Phi]} dt_{j},\quad i=1,\dots,k,\\
\tilde t_{q}=t_{q},\quad q=1,2,\dots,r,\quad
q\neq s_p\quad\mbox{for any}\quad p=1,\dots,k.
\end{array}
\end{equation}

In analogy with Proposition~\ref{eomp} we can prove the following result.
\begin{prop}
\label{eomp1} 
Suppose that  $\{H_{i}, H_{j}\}=0$
for all $i,j=1,\dots,r$ and $r\leq \corank P+(1/2)\rank P$, and
consider
the equations of motion (\ref{eomk1}) for $H_{i}$, $i=1,\dots,r$, restricted
onto $N_{\tilde\alpha}$.

Then the transformation
(\ref{rct}) is well defined on these restricted equations of motion
and sends them into the equations of
motion (\ref{eomk2}) for $\tilde H_{i}$, $i=1,\dots,r$,
restricted onto
$\tilde N_\alpha$.
\end{prop}

Note that the transformations from Propositions~\ref{eomp} and \ref{eomp1}
do not change the dynamical variables $x$. In particular,
under the assumptions of Proposition~\ref{eomp} for any given $i$ from 1 to $k$
the trajectories of the dynamical system
associated with $H_{s_i}$
are identical to those of the dynamical system
associated with $\tilde H_{s_i}$, if we consider the trajectories
as {\em non-parametrized} curves. In other words, the transformation
(\ref{rt0}) amounts to the reparametrization of the times associated
with $H_{s_j}$ for all $j=1,\dots,k$. Notice, however, 
that in general the reparametrization in question is {\em different} for
different trajectories, as one can readily infer from (\ref{rt0}).
\looseness=-1

As a final remark note that it could be interesting to compare the
above reparametrization results with those arising in the theory of
projectively equivalent metrics \cite{bm, top}.



\section{Canonical Poisson structure}


In this section we tacitly assume that $\tilde H_i$, $i=1,\dots,r$,
are related to $H_i$, $i=1,\dots,r$, through the $k$-parameter St\"ackel transform
(\ref{dual1}), (\ref{hi}) generated by $H_{s_1},\dots, H_{s_k}$.
We further assume that $M=\mathbb{R}^{2n}$, $P$~is
a canonical Poisson structure on $M$, and
$\lambda_{i}$, $\mu_{i}$, $i=1,\dots,n$, are the Darboux coordinates for $P$,
i.e.,
$\{\lambda_{i},\mu_{j}\}=\delta_{ij}$.
Let $\boldsymbol{\lambda}=(\lambda_{1},\dots,\lambda_{n})$ and
$\boldsymbol{\mu}=(\mu_{1},\dots,\mu_{n})$.
%
Then the Hamilton--Jacobi equations for $H_{i}$ and $\tilde H_{i}$ have a
common solution, cf.\ \cite{bkm}. Namely, we have the following generalization of
the results of \cite{bkm} to the case of multiparameter generalized St\"ackel transform:

\begin{prop}
\label{shjetr} 

Suppose that $\{H_{s_i}, H_{s_j}\}=0$
for all $i,j=1,\dots,k$. Let $S=S(\boldsymbol{\lambda},\alpha_1,\dots,\alpha_k, E_{s_1},\dots,E_{s_k},\allowbreak
a_{1},\dots,a_{n-k})$, where
$a_{i}$ are arbitrary constants, be a complete integral of the stationary
Hamilton--Jacobi equation for the Hamiltonians
$H_{s_i}=H_{s_i}(\boldsymbol{\lambda}, \boldsymbol{\mu},\alpha_1,\dots,\alpha_k)$,
\[
H_{s_i}(\boldsymbol{\lambda}, \partial S/\partial\boldsymbol{\lambda
}, \alpha_1,\dots,\alpha_k)=E_{s_i},\quad i=1,\dots,k.
\]

If we set $E_{s_i}=\tilde\alpha_i$ and $\alpha_i=\tilde E_{s_i}$ for all $i=1,\dots,k$ then
$S$ also is a complete
integral of the stationary Hamilton--Jacobi equation for the Hamiltonians
$\tilde H_{s_i}=\tilde H_{s_i}(\boldsymbol{\lambda},
\boldsymbol{\mu},\tilde{\alpha}_1,\dots\tilde{\alpha}_k)$,
\[
\tilde H_{s_i}(\boldsymbol{\lambda}, \partial S/\partial
\boldsymbol{\lambda}, \tilde{\alpha}_1,\dots\tilde{\alpha}_k)=\tilde E_{s_i}.
\]

Further assume that $r\leq n$,
and $\{H_{i},H_{j}\}=0$, $i,j=1,\dots,r$, and
\begin{equation}
\label{ci0}S=S(\boldsymbol{\lambda},\alpha_1,\dots,\alpha_k, E_{1},\dots,E_{r}, a_{1}%
,\dots,a_{n-r})
\end{equation}
where $a_{i}$ are arbitrary constants, be a complete integral for the system
of stationary Hamilton--Jacobi equations
\[
H_{i}(\boldsymbol{\lambda}, \partial S/\partial\boldsymbol{\lambda
},\alpha_1,\dots,\alpha_k)=E_{i}, \quad i=1,\dots,r.
\]

If we set
\[
\alpha_j=\tilde E_{s_j},\quad E_{s_j}=\tilde\alpha_j,\quad j=1,\dots,k,\quad\mbox{and}\quad
E_{i}=\tilde E_{i},\quad
i=1,\dots,r,\quad i\neq s_p\quad\mbox{for all}\quad p=1,\dots,k,
\]
then $S$ (\ref{ci0}) is also a complete integral for the system
\[
\tilde H_{i}(\boldsymbol{\lambda}, \partial S/\partial
\boldsymbol{\lambda},\tilde{\alpha}_1,\dots,\tilde{\alpha}_k)=\tilde E_{i}, \quad i=1,\dots,r.
\]
\end{prop}

This result suggests that the multiparametric generalized St\"ackel transform
potentially is a very powerful tool for solving the Hamilton--Jacobi
equations (and hence the equations of motion) for Hamiltonian
dynamical systems. Indeed, if we can solve the stationary
Hamilton--Jacobi equations for the original Hamiltonians $H_i$,
then by Proposition~\ref{shjetr}
we can do this for the transformed Hamiltonians $\tilde
H_i$ as well, and vice versa.\looseness=-1

As for the equations of motion, 
in addition to general
Propositions~\ref{eomp} and \ref{eomp1}, a somewhat more explicit result
can be obtained by straightforward computation:
\begin{cor}
\label{reoms}
Suppose that $r=n$, $\{H_i,H_j\}=0$ for all $i,j=1,\dots,n$,
$\partial^2 H_{i}/\partial\alpha_j\partial\boldsymbol{\mu}=0$ for all
$i=1,\dots,n$ and all $j=1,\dots,k$,
and that $\lambda_{j}$, $j=1,\dots,n$, can be chosen as local coordinates on the
Lagrangian submanifold $N_{E}=\{ (\boldsymbol{\lambda}, \boldsymbol{\mu})\in
M| H_{i}(\boldsymbol{\lambda}, \boldsymbol{\mu},\alpha_1,\dots,\alpha_k)=E_{i}, \quad
i=1,\dots,n\}$ (in other words, the system $H_{i}(\boldsymbol{\lambda},
\boldsymbol{\mu},\alpha_1,\dots,\alpha_k)=E_{i}$, $i=1,\dots,n$, can be solved for $\boldsymbol{\mu}$),
and that we have
\begin{equation}
\label{par0}\hspace*{-2.7mm}\alpha_j=\tilde E_{s_j},\quad\!\! E_{s_j}=\tilde\alpha_j,\quad\!\! j=1,\dots,k,
\quad\!\!\mbox{and}\quad\!\!
E_{i}=\tilde E_{i},\quad\!\!
i=1,\dots,n,\quad\!\! i\neq s_p\quad\mbox{for all}\quad\!\! p=1,\dots,k.
\end{equation}

Then the reciprocal transformation (\ref{rct}) turns the system
\begin{equation}
\label{lt1}d\boldsymbol{\lambda}/ dt_{i}=(\partial H_{i}/\partial
\boldsymbol{\mu})|_{N_{E}},\quad i=1,\dots,n,
\end{equation}
into
\begin{equation}
\label{lt2}d\boldsymbol{\lambda}/d\tilde t_{i} =(\partial\tilde H_{i}%
/\partial\boldsymbol{\mu})|_{\tilde N_{\tilde E}},\quad i=1,\dots,n,
\end{equation}
where $\tilde N_{\tilde E}=\{ (\boldsymbol{\lambda}, \boldsymbol{\mu})\in M|
\tilde H_{i}(\boldsymbol{\lambda}, \boldsymbol{\mu},\tilde{\alpha}_1,\dots,
\tilde{\alpha}_k)=\tilde
E_{i}, \quad i=1,\dots,n\}$.
\end{cor}
Recall that $N_E$ and $N_{\tilde{E}}$ in fact represent {\em the same}
Lagrangian submanifold of $M$, cf.\ the remark after Proposition~\ref{eomp}.

For instance, if we have $k=1$, $\alpha_1\equiv\alpha$, $s_1=s$, and take
\begin{equation}
\label{nat}H_{i}=\displaystyle\frac12(\boldsymbol{\mu}, G_{i}%
(\boldsymbol{\lambda})\boldsymbol{\mu}) +V_{i}(\boldsymbol{\lambda}) +\alpha
W_{i}(\boldsymbol{\lambda}), \quad i=1,\dots,n,
\end{equation}
where $(\cdot,\cdot)$ stands for the standard scalar product in $\mathbb{R}^{n}$ and
$G_{i} (\boldsymbol{\lambda})$ are $n\times n$ matrices, then the system
(\ref{lt1}) reads
\begin{equation}
\label{lm1}d\boldsymbol{\lambda}/ dt_{i}=G_{i}%
(\boldsymbol{\lambda })\boldsymbol{M},
\end{equation}
where $\boldsymbol{\mu}=\boldsymbol{M}(\boldsymbol{\lambda},\alpha,E_{1}%
,\dots,E_{n})$ is a general solution of the system $H_{i}(\alpha
,\boldsymbol{\lambda},\boldsymbol{\mu})=E_{i}$, $i=1,\dots,n$.

If we eliminate $\boldsymbol{M}$ from (\ref{lm1}) then we obtain the
\emph{dispersionless Killing systems} (cf.\ \cite{m,blasak,f1,f2})
\begin{equation}
\label{ki}\boldsymbol{\lambda}_{t_{i}}=G_{i} (G_{s})^{-1}%
\boldsymbol{\lambda }_{t_{s}},\quad i=1,2,\dots,s-1,s+1,\dots,n,
\end{equation}
and
the reciprocal transformation (\ref{rct}), which in our case reads
\[
d\tilde t_{s}=-\sum\limits_{i=1}^{n} W_{i}(\boldsymbol{\lambda})d t_{i}%
,\qquad\tilde t_{i}=t_{i},\quad i\neq s, 
\]
turns (\ref{ki}) into
\begin{equation}
\label{ki2}\boldsymbol{\lambda}_{\tilde t_{i}}=\tilde G_{i} (\tilde
G_{s})^{-1}\boldsymbol{\lambda}_{\tilde t_{s}}, \quad i=1,2,\dots
,s-1,s+1,\dots,n,
\end{equation}
where the quantities
$\tilde G_{s}=-G_{s}/W_{s}$ and $\tilde G_{i}=G_{i}-W_{i} G_{s}/W_{s}$,
$i=1,2,\dots,s-1,\allowbreak s+1,\dots,n$, are related to the Hamiltonians
\begin{equation}
\tilde H_{i}=\displaystyle\frac12(\boldsymbol{\mu},\tilde G_{i}%
(\boldsymbol{\lambda})\boldsymbol{\mu}) +\tilde V_{i}(\boldsymbol{\lambda})
+\tilde\alpha\tilde W_{i}(\boldsymbol{\lambda}), \quad i=1,\dots,n,
\end{equation}
which are St\"ackel-equivalent to $H_i$, $i=1,\dots,n$.


We can now apply Proposition~\ref{shjetr} in order to obtain the
solutions of equations of motion (\ref{lt1}) and
(\ref{lt2}) as follows: 
\begin{cor}
\label{soleom} Under the assumptions of Corollary~\ref{reoms}, suppose that
\begin{equation}
\label{ci} S=S(\boldsymbol{\lambda},\alpha_1,\dots,\alpha_k, E_{1},\dots,E_{n})
\end{equation}
is a complete integral for the system of stationary Hamilton--Jacobi
equations
\[
H_{i}(\boldsymbol{\lambda}, \partial S/\partial\boldsymbol{\lambda
},\alpha_1,\dots,\alpha_k)=E_{i}, \quad i=1,\dots,n.
\]

Then a general solution of (\ref{lt1}) for $i=d$ can be written in implicit
form as
\begin{equation}
\label{is1}\partial S/\partial E_{j}=\delta_{jd} t_{d}+b_{j},\quad
j=1,\dots,n,
\end{equation}
where $b_{j}$ are arbitrary constants, and by virtue of (\ref{par0}) a general
solution of (\ref{lt2}) for $i=d$ can be written in implicit form as
\begin{equation}
\label{is2}\partial S/\partial\tilde E_{j}=\delta_{jd} \tilde t_{d}+ b_{j},\quad j=1,\dots,n.
\end{equation}
\end{cor}

Comparing (\ref{is1}) and (\ref{is2}) and using (\ref{par0}) we readily see
that, in perfect agreement with (\ref{rct}),
$t_{i}=\tilde t_{i}$ for $i\neq s_1,\dots,s_k$, but
$t_{s_j}=\partial S/\partial E_{s_j}- b_{s_j}=\partial
S/\partial\tilde{\alpha}_j-b_{s_j}$ while $\tilde t_{s_j}=\partial
S/\partial\tilde E_{s_j}-b_{s_j}=\partial S/\partial
\alpha_j-b_{s_j}$. Thus, the above approach does not yield an
\emph{explicit} formula expressing $\tilde t_{s_j}$ as functions
of $\boldsymbol{\lambda}, \boldsymbol{\mu}$, and $t_{s_i}$.
\looseness=-1

In order to find a complete integral (\ref{ci}) we can use
separation of variables as follows (see e.g.\ \cite{Sklyanin,
mac2005} and references therein).
Under the assumptions of Corollary~\ref{soleom} suppose that
$\lambda_{i}$, $\mu_{i}$, $i=1,\dots,n$, are separation coordinates
for the Hamiltonians $H_{i}$, $i=1,\dots,n$, that is, the system of
equations
$H_{i}(\boldsymbol{\lambda}, \boldsymbol{\mu},
\alpha_1,\dots,\alpha_k)=E_{i}$, $i=1,\dots,n$,
is equivalent to the following one:\looseness=-1
\begin{equation}
\label{sr}\varphi_{i}(\lambda_{i},\mu_{i},\alpha_1,\dots,\alpha_k,
E_{1},\dots,E_{n})=0, \quad i=1,\dots,n, 
\end{equation}\nopagebreak[4]
which is nothing but the set of the separation relations\footnote{Note
that the separation relations involving parameters
appear, in a rather different context, in
the paper \cite{ts3} where they are employed for
the construction of separation variables.}
on the Lagrangian submanifold $N_{E}$.

On the other hand, under the identification (\ref{par0}) the
system (\ref{sr}) is equivalent to
\begin{equation}
\label{shje2}\tilde H_{i}(\boldsymbol{\lambda},
\boldsymbol{\mu},\tilde\alpha_1,\dots,\tilde\alpha_k)=\tilde
E_{i}, \quad i=1,\dots,n.
\end{equation}
Thus, the St\"ackel-equivalent $n$-tuples of Hamiltonians share the separation relations
(\ref{sr}) provided (\ref{par0}) holds.\looseness=-1

Consider the system of stationary Hamilton--Jacobi equations for $H_{i}$
\begin{equation}
\label{shje}H_{i}(\boldsymbol{\lambda}, \partial S/\partial
\boldsymbol{\lambda},\alpha_1,\dots,\alpha_k)=E_{i}, \quad i=1,\dots,n.
\end{equation}
By the above, (\ref{shje}) is equivalent to the system
\begin{equation}
\label{sshje}\varphi_{i}(\lambda_{i},\partial S/\partial\lambda_{i},\alpha_1,\dots,\alpha_k,
E_{1},\dots,E_{n})=0, \quad i=1,\dots,n.
\end{equation}

Suppose that (\ref{sr}) can be solved for $\mu_{i}$, $i=1,\dots,n$:
\[
\mu_{i}=M_{i}(\lambda_{i},\alpha_1,\dots,\alpha_k,E_{1},\dots,E_{n}), \quad i=1,\dots,n.
\]

Then there exists a separated complete integral of (\ref{sshje}), and hence of
(\ref{shje}), of the form (cf.\ e.g.\ \cite{mac2005}) 
\begin{equation}
\label{cis}S=\sum\limits_{l=1}^{n} \int M_{l}(\lambda_{l},\alpha_1,\dots,\alpha_k,E_{1}%
,\dots,E_{n})d\lambda_{l},
\end{equation}
and general solutions for (\ref{lt1}) and (\ref{lt2}) can be found using the
method of Corollary~\ref{soleom}.\looseness=-1

In this case the formulas (\ref{is1}) take the form
\begin{equation}\label{is1a}
\sum\limits_{i=1}^{n} \int(\partial M_{i}(\lambda_{i},\alpha_1,\dots,\alpha_k,E_{1},\dots
,E_{n})/\partial E_{j})d\lambda_{i} =\delta_{jd} t_{d}+b_{j},\quad
j=1,\dots,n,
\end{equation}
and expressing $\lambda_i$ as functions of $t_d$ from (\ref{is1a}) 
is nothing but an instance of the Jacobi inversion problem.\looseness=-1

In particular,  for $d=s_i$ we have
\[
\tilde t_{s_i}+b_{s_i}=\partial S/\partial\tilde E_{s_i}=\partial S/\partial\alpha_i
=\sum\limits_{l=1}^{n} \int(\partial M_{l}(\lambda_{l},\alpha_1,\dots,\alpha_k,E_{1}%
,\dots,E_{n})/\partial\alpha_i)d\lambda_{l}, \quad i=1,\dots,k.
\]


\section{Multiparameter generalized St\"ackel transform\\ and deformations of separation curves}

Under the assumptions of Corollary~\ref{reoms}, suppose that
$\lambda_{i}$, $\mu_{i}$, $i=1,\dots,n$, are \emph{separation
coordinates} for the $n$-tuple of commuting Hamiltonians $H_{i}$,
$i=1,\dots,n$. Then the Lagrangian submanifold $N_{E}$ is defined
by $n$ separation relations (\ref{sr}). Further assume that all
functions $\varphi_{i}$
are identical:
\begin{equation}
\label{sk}\varphi_{i}=\varphi(\lambda_{i},\mu_{i},\alpha_1,\dots,\alpha_k,
E_{1},\dots,E_{n}), \quad i=1,\dots,n.
\end{equation}
Then relations (\ref{sr}) mean that 
the points
$(\lambda_{i},\mu_{i})$, $i=1,\dots,n$,
belong to the {\em separation curve} \cite{Sklyanin, mac2005}
\begin{equation}
\label{sk1}\varphi(\lambda,\mu,\alpha_1,\dots,\alpha_k,
E_{1},\dots,E_{n})=0.
\end{equation}
\looseness=-1

If the relations
\[
\varphi(\lambda_{i},\mu_{i},\alpha_1,\dots,\alpha_k,H_{1},\dots,H_{n})=0,
\quad i=1,\dots,n,
\]
uniquely determine the Hamiltonians $H_{i}$ for $i=1,\dots,n$, then
for the sake of brevity we shall say that $H_{i}$ for $i=1,\dots,n$
have the {\em separation curve}
\begin{equation}\label{sepcu}
\varphi(\lambda,\mu,\alpha_1,\dots,\alpha_k, H_{1},\dots,H_{n})=0.
\end{equation}
Fixing values of all Hamiltonians $H_{i}=E_{i}$, $i=1,\dots,n$,
picks a particular Lagrangian submanifold from the Lagrangian
foliation. It is also clear that the St\"ackel-equivalent
$n$-tuples of the Hamiltonians $H_{i}$, $i=1,\dots,n$, and $\tilde H_{i}$,
$i=1,\dots,n$, share the separation curve (\ref{sepcu})
provided (\ref{dual1}) and (\ref{dual2}) hold.

In the rest of this section we shall deal with a special class
of separation curves of the form (cf.\ e.g.\ \cite{mac2005} and references therein)
\begin{equation}
\label{kska0}\sum\limits_{j=1}^n H_{j}\lambda^{\beta_j}=\psi(\lambda,\mu),
\end{equation}
where $\beta_j$ are arbitrary pairwise distinct non-negative integers,
$\beta_1>\beta_2>\cdots>\beta_n$. In fact one always can impose
the normalization $\beta_n=0$ by dividing the left- and right-hand side
of (\ref{kska0}) by $\lambda^{\beta_n}$ if necessary, but we shall not impose this
normalization in the present paper.


For a given $n$, each class of systems (\ref{kska0}) is
labelled by a sequence $(\beta_{1},\dots,\beta_{n})$ while a
particular system from a class is given by a particular choice of
$\psi(\lambda,\mu)$. In particular, the choice
$\psi(\lambda,\mu)=\frac{1}{2}f(\lambda)\mu^{2}+\gamma(\lambda)$
yields the well-known classical St\"ackel systems.
All these systems admit the separation of variables in the same
coordinates $(\lambda_{i},\mu_{i})$ by construction.

We shall refer to the class with the separation curve
\begin{equation}\label{bsk0a}
\sum\limits_{j=1}^n H_j \lambda^{n-j}=\psi(\lambda,\mu)
\end{equation}
as to the {\em seed class}. Note that if
$\psi(\lambda,\mu)=\frac{1}{2}f(\lambda)\mu^{2}+\gamma(\lambda)$ we obtain
precisely the Benenti class of St\"ackel systems
\cite{ben93,ben97}. The seed class is a rather general one:
it includes the majority of known integrable systems
with natural Hamiltonians
\cite{mac2005}.

It turns out that, roughly speaking,
the $n$-tuple of Hamiltonians having
the general separation curve (\ref{kska0})
can be related via a suitably
chosen generalized multiparameter St\"ackel transform
to an $n$-tuple of Hamiltonians having the separation curve
(\ref{bsk0a}) from the seed class.
The exact picture is a bit more involved, as
in fact we need to consider
the {\em deformations} of the curves in question.

Define first an operator $R_k^f$ that acts
as follows:
\[
R_k^f(F)=F+ f \lambda^k -(\lambda^k/k!)(\p^k F/\p\lambda^k)|_{\lambda=0}.
\]
For instance, we have
\[
R_k^f \left(\sum\limits_{j=0}^s a_j\lambda^j\right)
=f \lambda^k+\sum_{j=0, j\neq k}^s a_j\lambda^j.
\]

Now let
\[
\begin{array}{l}
F_0=\sum\limits_{j=1}^n H_{j}\lambda^{n-j}\quad\mbox{and}\quad
\tilde F_0=\sum\limits_{j=1}^n \tilde H_{j}\lambda^{n-j}.
\end{array}
\]

For any integer $m$ define \cite{mac2005} the so-called
basic separable potentials $V_j^{(m)}$ by means of the relations
\begin{equation}\label{sp}
\lambda^m+ \sum\limits_{j=1}^n V_j^{(m)}\lambda^{n-j}=0
\end{equation}
that must hold for $\lambda=\lambda_i$, $i=1,\dots,n$.

Under the assumptions of Corollary~\ref{reoms}, consider an
$n$-tuple of commuting Hamiltonians of the form\looseness=-1
\begin{equation}\label{bend}
H_i=H_i^{(0)}+\sum\limits_{j=1}^k\alpha_j V_i^{(\gamma_j)},
\end{equation}\nopagebreak
where $\gamma_j$, $j=1,\dots,k$, are pairwise distinct integers.

Suppose that the Hamiltonians (\ref{bend}) have the separation
curve of the form 
\begin{equation}\label{bsk}
\sum\limits_{j=1}^k \alpha_{j}\lambda^{\gamma_j}+F_0=\psi(\lambda,
\mu),
\end{equation}
\nopagebreak[4]
where $\gamma_j>n-1$ for all $j=1,\dots,k$,
and $\gamma_i\neq \gamma_j$ if $i\neq j$
for all $i,j=1,\dots,k$.\looseness=-1

Now pick $k\leq n$ distinct numbers $s_i\in\{1,\dots,n\}$
and define the Hamiltonians $\tilde H_i$ by means of the following
separation curve
\begin{equation}
\sum\limits_{j=1}^k \tilde H_{s_j}\lambda^{\gamma_j}+
R_{n-s_1}^{\tilde\alpha_1}\cdots R_{n-s_k}^{\tilde\alpha_k}(\tilde
F_0)=\psi(\lambda, \mu). \label{ksk}
\end{equation}
This means that $\tilde H_i$ are the solutions of the system of
linear algebraic equations obtained from (\ref{ksk}) upon
substituting $\lambda_i$ for $\lambda$ and $\mu_i$ for $\mu$ into
(\ref{ksk}) for $i=1,\dots,n$.

\begin{prop}\label{dscp} Under the above assumptions
the $n$-tuple of Hamiltonians $\tilde H_i$, $i=1,\dots,n$, is
St\"ackel-equivalent to $H_i$, $i=1,\dots,n$.

The $n$-parameter generalized St\"ackel transform relating $\tilde H_i$, $i=1,\dots,n$
to $H_i$, $i=1,\dots,n$ reads as follows:
\begin{equation}\label{dual1lina}
\tilde H_{s_i}=\det B_i/\det B,
\end{equation}
where
\[
B=\left|\!\left|\begin{array}{ccc}
V_{s_1}^{(\gamma_1)} & \cdots & V_{s_1}^{(\gamma_k)}\\
\vdots & \ddots & \vdots\\
V_{s_k}^{(\gamma_1)} & \cdots & V_{s_k}^{(\gamma_k)}
\end{array}\right|\!\right|
\]
is a $k\times k$ matrix, and $B_i$ are obtained from $B$ by
replacing $V_{s_j}^{(\gamma_{i})}$ by
$H_{s_j}^{(0)}-\tilde\alpha_j$ for all $j=1,\dots,k$;
\begin{equation}
\label{hilina}%
\tilde H_{i}
=H_{i}^{(0)} +\sum\limits_{j=1}^k \tilde H_{s_j}
V_{i}^{(\gamma_j)}, \qquad i=1,\dots,r, \quad i\neq
s_j\quad\mbox{for}\quad\! j=1,\dots,k,
\end{equation}
where $\tilde H_{s_i}$ are given by (\ref{dual1lina}).
\end{prop}

{\em Proof.} First of all, note that the above formulas for
$\tilde H_i$ indeed constitute the St\"ackel transform, as
Eq.(\ref{dual1lina}) is readily seen to imply the relations of the
type (\ref{dual1}), namely
\begin{equation}
\label{dual1linb} H_{s_i}^{(0)}+\sum\limits_{j=1}^k \tilde H_{s_j}
V_{s_i}^{(\gamma_j)}=\tilde{\alpha}_i,\quad i=1,\dots,k,
\end{equation}
cf.\ the discussion after (\ref{dual1lin0}).

Now we only have to prove that the Hamiltonians $\tilde H_i$ defined by
(\ref{dual1lina}) and (\ref{hilina}) have the separation curve
(\ref{ksk}). As we have already mentioned above, the
St\"ackel-equivalent $n$-tuples of separable commuting
Hamiltonians share the separation relations provided (\ref{par0})
holds. Therefore, in order to prove our claim it suffices to show
that the separation curves (\ref{bsk}) and (\ref{ksk}) can be
identified by virtue of (\ref{dual1linb}).
\looseness=-1

Indeed, upon plugging into (\ref{bsk}) the relations
\begin{equation}
\label{sp1}
\lambda^{\gamma_j}=-\sum\limits_{p=1}^n V_p^{(\gamma_j)}\lambda^{n-p},\quad j=1,\dots,k,
\end{equation}
that follow from (\ref{sp}), collecting the coefficients at the powers of $\lambda$,
and taking into account (\ref{bend}),
the separation curve (\ref{bsk}) can be rewritten as
\begin{equation}
\label{bsk0}\sum\limits_{j=1}^n H_{j}^{(0)}\lambda^{n-j}=\psi(\lambda,\mu).
\end{equation}

On the other hand, plugging (\ref{sp1}) into (\ref{ksk}) and proceeding
in a similar fashion as above, we obtain
\begin{equation}
-\sum\limits_{p=1}^n \left(\sum\limits_{j=1}^k \tilde
H_{s_j}V_p^{(\gamma_j)}\right)\lambda^{n-p}+
R_{n-s_1}^{\tilde\alpha_1}\cdots R_{n-s_k}^{\tilde\alpha_k}(\tilde
F_0)=\psi(\lambda, \mu). \label{ksk1}
\end{equation}
By virtue of relations (\ref{dual1linb}), which can be further
rewritten as
\[
H_{s_i}^{(0)}=-\sum\limits_{j=1}^k\tilde H_{s_j} V_i^{(\gamma_j)}+\tilde\alpha_i,\quad i=1,\dots,k,
\]
along with (\ref{hilina}),
we find that the curves (\ref{ksk1}) and (\ref{bsk0}) are indeed identical, and
hence so are the curves (\ref{ksk}) and (\ref{bsk}). $\square$\looseness=-1

\begin{remk} 
In fact the above argument can be inverted,
that is, we can obtain the relations (\ref{dual1linb}) (and hence
(\ref{dual1lina})) and (\ref{hilina}) by requiring the curves
(\ref{bsk}) and (\ref{ksk}) to coincide and comparing the
coefficients at the powers of $\lambda$ on the left-hand sides of
these curves, or equivalently (by virtue of (\ref{sp})), of
(\ref{ksk1}) and (\ref{bsk0}).\looseness=-1
\end{remk}

\begin{prop}\label{dscpi} 
The inverse of the $k$-parameter generalized St\"ackel transform
(\ref{dual1lina}), (\ref{hilina})
has the form:
\begin{equation}\label{dual1linai}
H_{s_i}=\det \tilde B_i/\det \tilde B,
\end{equation}
where
\[
\tilde B=\left|\!\left|\begin{array}{ccc}
\tilde V_{s_1}^{(n-s_1)} & \cdots & \tilde V_{s_1}^{(n-s_k)}\\
\vdots & \ddots & \vdots\\
\tilde V_{s_k}^{(n-s_1)} & \cdots & \tilde V_{s_k}^{(n-s_k)}
\end{array}\right|\!\right|
\]
is a $k\times k$ matrix, and $\tilde B_i$ are obtained from $\tilde B$ by
replacing $\tilde V_{s_j}^{(n-s_{i})}$ by
$\tilde H_{s_j}^{(0)}-\alpha_j$ for all $j=1,\dots,k$;
\begin{equation}
\label{hilinai}%
H_{i}
=\tilde H_{i}^{(0)} +\sum\limits_{j=1}^k H_{s_j} \tilde
V_{i}^{(n-s_j)}, \qquad i=1,\dots,r, \quad i\neq
s_j\quad\mbox{for}\quad\! j=1,\dots,k,
\end{equation}
where $H_{s_i}$ are given by (\ref{dual1linai}) and $\tilde V_{j}^{(m)}$ are
{\em deformed separable potentials}
defined for all integer $m$ {\em except} $m=\gamma_i$, $i=1,\dots,k$,
by means of the relations
\begin{equation}\label{spi}
\lambda^m+\sum\limits_{j=1}^k \tilde V_{s_j}^{(m)}\lambda^{\gamma_j}
+\sum\limits_{p=1, p\neq s_1,\dots,s_k}^n \tilde V_p^{(m)}\lambda^{n-p}=0
\end{equation}
that must hold for $\lambda=\lambda_i$, $i=1,\dots,n$.
\end{prop}

The proof of this result is readily obtained from that of Proposition~\ref{dscp} using the fact
that the inverse of the $n$-parameter generalized St\"ackel transform (\ref{dual1lina}), (\ref{hilina})
is nothing but the dual of the latter (see Section~\ref{mgstdefsect} for the definition of duality).

As a final remark, note that
upon setting the parameters $\alpha_i$ and $\tilde\alpha_i$
to zero for all $i=1,\dots,k$ the formulas (\ref{dual1lina}) and (\ref{hilina})
indeed relate the Hamiltonians $H_i$ with the separation curve (\ref{bsk0a})
and the Hamiltonians $\tilde H_i$ with the separation curve (\ref{kska0}).
In this case we essentially  recover
the formulas from \cite{mac2005} relating the Hamiltonians from the seed class and
from the so-called $k$-hole deformation thereof (in our language, the deformed systems are
precisely those having the separation curve (\ref{kska0}))
up to a suitable renumeration of the Hamiltonians $\tilde H_i$.
\looseness=-1

\section{Examples}
As a simple illustration of the above results,
consider the Hamiltonian systems on a four-dimensional phase space
$M=\mathbb{R}^4$ with the coordinates $(p_{1},p_{2},q_{1},q_{2})$
and canonical Poisson structure.

For our first example let $k=1$, $r=2$, $s_1=2$,
$\alpha_1\equiv\alpha$ and $\tilde\alpha_1\equiv\tilde\alpha$.
Consider the Hamiltonian
\[
H_1=\frac{1}{2}p_{1}^{2}+\frac{1}{2}p_{2}^{2}+ \frac{\alpha
(q_1^2-q_2^2)}{q_2}p_2-2\alpha^2 q_1^2,
\]
which is Liouville integrable because it Poisson commutes with
\[
H_{2} =\frac{q_1 p_2-q_2 p_1-2\alpha q_1 q_2}{p_2}.
\]
The above pair of commuting Hamiltonians was found by analogy
with one of the models from \cite{hiet2}.

The relation (\ref{dual1}) in this case takes the form
\[
\frac{q_1 p_2-q_2 p_1-2\tilde H_2 q_1 q_2}{p_2}=\tilde\alpha,
\]
whence
\[
\tilde H_2=\frac{q_1 p_2-q_2 p_1-\tilde\alpha p_2}{2 q_1 q_2},
\]
and therefore by virtue of (\ref{hi}) we have
\[
\tilde H_1=\frac{q_1^2+q_2^2-2 \tilde\alpha q_1}{2 q_1 q_2}p_1 p_2
+\frac{\tilde\alpha(q_1^2-\tilde\alpha q_1+q_2^2)}{2 q_1
q_2^2}p_2^2.
\]
By Proposition~\ref{trp}, ii) the relation $\{H_1,H_2\}=0$ implies
$\{\tilde H_1,\tilde H_2\}=0$, so $\tilde H_1$ is Liouville integrable just like $H_1$.
Interestingly enough,
in this example the generalized St\"ackel
transform sends the Hamiltonian $H_1$ into a natural
{\em geodesic} Hamiltonian $\tilde H_1$, but the metric associated
with $\tilde H_1$ is not flat and, moreover, has nonconstant scalar curvature
unlike the metric associated with $H_1$.
\looseness=-1

By Proposition~\ref{eomp}
the reciprocal transformation
\[
\tilde t_1=t_1,\quad
d\tilde t_2=\left(-2 q_1 p_1+\frac{(q_1^2-2\tilde\alpha q_1+q_2^2)p_2}{q_2}\right) dt_1+\frac{2q_1 q_2}{p_2} dt_2
\]
takes
the equations of motion for $H_1$ and $H_2$, with the respective
evolution parameters $t_1$ and $t_2$, restricted onto the common level surface
$N_{\tilde\alpha}=\{x\in\mathbb{R}^4|H_1(x,\alpha_1,\alpha_2)=\tilde\alpha_1,
H_2(x,\alpha_1,\alpha_2)=\tilde\alpha_2\}$ into the
equations of motion for $\tilde H_1$ and $\tilde H_2$, with the respective
evolution parameters $\tilde t_1$ and $\tilde t_2$, restricted onto the common level surface
$\tilde N_{\alpha}=\{x\in\mathbb{R}^4|\tilde H_1(x,\tilde\alpha_1,\tilde\alpha_2)=\alpha_1,
\tilde H_2(x,\tilde\alpha_1,\tilde\alpha_2)=\alpha_2\}$. It is easily seen that
$\tilde N_{\alpha}$ and $N_{\tilde\alpha}$ indeed represent the same submanifold
of $\mathbb{R}^4$.

For the second example we set $k=r=2$ and consider the
(extended) H\'enon--Heiles system with the Hamiltonian
\[
H_{1} =\frac{1}{2}p_{1}^{2}+
\frac{1}{2}p_{2}^{2}-\alpha_1\left(q_1^3+\frac{q_1  q_2^2}{2}\right)
-\alpha_2 q_1,
\]
which Poisson commutes with
\[
H_{2} =\frac{1}{2}q_{2}p_{1}p_{2}-\frac{1}{2}q_{1}p_{2}^{2}-
\alpha_1\left(\frac{q_2^4}{16}+\frac{q_1^2 q_2^2}{4}\right)
-\alpha_2\frac{q_2^2}{4}.
\]

The separation curve for the system in question belongs to the seed class and reads
\begin{equation}\label{schh}
\alpha_1\lambda^4+\alpha_2\lambda^2+H_1\lambda+H_2=\lambda\mu^2/2.
\end{equation}
The separation coordinates $(\lambda_i,\mu_i)$, $i=1,2$, are
related to $p$'s and $q$'s by the formulas
\[
\begin{array}
[c]{l}%
q_{1}=\lambda_{1}+\lambda_{2},\qquad q_{2}=2\sqrt{-\lambda_{1}\lambda_{2}%
},\\[5mm]%
\displaystyle p_{1}=\frac{\lambda_{1}\mu_{1}}{\lambda_{1}-\lambda_{2}}%
+\frac{\lambda_{2}\mu_{2}}{\lambda_{2}-\lambda_{1}},\qquad p_{2}%
=\sqrt{-\lambda_{1}\lambda_{2}}\left(  \frac{\mu_{1}}{\lambda_{1}-\lambda_{2}%
}+\frac{\mu_{2}}{\lambda_{2}-\lambda_{1}}\right).
\end{array}
\]

Let 
$s_1=1$, $s_2=2$, $k=r=2$.
Then (\ref{bend}) and (\ref{dual1lina}) yield
the following deformation of $H_1$ and $H_2$:
\[
\begin{array}{l}
\displaystyle \tilde H_{1} =\frac{2}{q_1 q_2^2} p_1^2-\frac{8}
{q_2^3}p_1 p_2-\frac{2(q_2^2+4 q_1^2)}{q_1 q_2^4}p_2^2
-\frac{4}{q_1 q_2^2}\tilde\alpha_1+\frac{16}{q_2^4}\tilde\alpha_2,
\\[5mm]\displaystyle
\tilde H_{2}=-\frac{4 q_1^2+q_2^2}{2 q_1 q_2^2}p_1^2
-\frac{4(q_2^2+2 q_1^2)}{q_2^3}p_1 p_2
+\frac{16 q_1^4+12 q_1^2 q_2^2+q_2^4}{q_1 q_2^4}p_2^2\\[4mm]\displaystyle\hspace*{7mm}
+\frac{(q_2^2+4 q_1^2)}{q_1 q_2^2}\alpha_1
-\frac{8(q_2^2+2 q_1^2)}{q_2^4}\alpha_2.
\end{array}
\]
The corresponding separation curve reads (see Proposition \ref{dscp})
\begin{equation}\label{schhd}
\tilde H_1 \lambda^4+\tilde H_2\lambda^2+\tilde\alpha_1\lambda+\tilde\alpha_2=\lambda\mu^2/2.
\end{equation}

Using Proposition~\ref{eomp} and proceeding
in analogy with the previous example we readily find
that the reciprocal transformation (\ref{rct})
for the equations of motion restricted onto the appropriate
Lagrangian manifolds in our case takes the form
\[
d\tilde t_1=\left(q_1^3+\frac{q_1  q_2^2}{2}\right) dt_1
+\left(\frac{q_2^4}{16}+\frac{q_1^2 q_2^2}{4}\right) dt_2,
\quad d\tilde t_2=q_1^2 dt_1+ \frac{q_2^2}{4} dt_2.
\]

\section*{Acknowledgments}

This research was supported in part by the Czech Grant Agency
(GA\v{C}R) under grant No.~201/04/0538, by the Ministry of
Education, Youth and Sports of the Czech Republic (M\v{S}MT \v{C}R)
under grant MSM 4781305904, by Silesian University in Opava under
grant IGS 10/2007, and by the Ministry of Science and Higher Education (MNiSW)
of the Republic of Poland under the research grant No.~N~N202~4049~33.
\looseness=-1

A.S. appreciates warm hospitality of the Institute of Physics of
the Adam Mickiewicz University in Pozna\'n, Poland, where the present work was
completed. A.S. would also like to thank Prof. A.V. Bolsinov,
Prof. W. Miller, Jr, and
Prof. S. Rauch-Wojciechowski for stimulating discussions and helpful comments.
The authors thank the referees for useful suggestions. \looseness=-1


\end{document}